\documentclass[a4paper]{jpconf}
\usepackage{graphicx}
\begin{document}
\title{Pulsations in carbon-atmosphere white dwarfs: A new chapter in
  white dwarf asteroseismology}
\author{G Fontaine$^{1}$, P Brassard$^1$, P Dufour$^2$, E M Green$^2$
  and J Liebert$^2$}

\address{$^1$ D\'epartement de Physique, Universit\'e de Montr\'eal,
  C.P. 6128, Succ. Centre-Ville, Montr\'eal, Qu\'ebec, Canada H3C 3J7\\
$^2$ Department of Astronomy and Steward Observatory, University of
  Arizona, 933 North Cherry Avenue, Tucson, AZ 85721, USA\\
}
\ead{fontaine@astro.umontreal.ca; brassard@astro.umontreal.ca;
  dufourpa@as.arizona.edu; bgreen@as.arizona.edu; liebert@as.arizona.edu} 

\begin{abstract}
We present some of the results of a survey aimed at exploring the
asteroseismological potential of the newly-discovered carbon-atmosphere
white dwarfs. We show that, in certains regions of parameter space,
carbon-atmosphere white dwarfs may drive low-order gravity modes. We
demonstrate that our theoretical results are consistent with the recent
exciting discovery of luminosity variations in SDSS J1426+5752 and some
null results obtained by a team of scientists at McDonald
Observatory. We also present follow-up photometric observations carried
out by ourselves at the Mount Bigelow 1.6-m telescope using the new
Mont4K camera. The results of follow-up spectroscopic observations at
the MMT are also briefly reported, including the surprising discovery
that SDSS J1426+5752 is not only a pulsating star but that it is also a
magnetic white dwarf with a surface field near 1.2 MG. The discovery of
$g$-mode pulsations in SDSS J1426+5752 is quite significant in itself as
it opens a fourth asteroseismological ``window'', after the GW Vir, V777
Her, and ZZ Ceti families, through which one may study white dwarfs. 
\end{abstract}

\section{Introduction}
Dufour et al. (2007) reported recently on the unexpected discovery of a
new type of white dwarfs, those with carbon-dominated atmospheres, also
known as Hot DQ stars. These are very rare, and Dufour et al. (2007)
wrote on the discovery of nine of those, out of a total of several
thousands white dwarfs already identified spectroscopically. The Hot DQ stars
were, in fact, uncovered within the framewok of the SDSS project, which
has led to many other surprises. Among other characteristics, the
C-atmosphere white dwarfs bunch together in a narrow range of effective
temperature, between about 18,000 K and 23,000 K (Dufour et al. 2008a).

It has been known for quite some time (see, e.g., Fontaine \& Van Horn 1976)
that models of C-atmosphere white dwarfs in this temperature range
possess superficial convection zones that bear strong similarities with
those found in H- and He-atmosphere stars. Given that the newly-found
C-atmosphere white dwarfs are sandwiched between the V777 Her and ZZ
Ceti instability strips, and given that convection plays a key role in
the excitation of pulsation modes in these pulsators, it follows that
the prospects of finding unstable models of Hot DQ stars looked good at
the outset.

\section{Stability survey of models of Hot DQ white dwarfs}

In an initial effort, three of us carried out an exploratory stability
survey of models of carbon-atmosphere white dwarfs using a full
nonadiabatic approach, and this was reported in Fontaine, Brassard, \&
Dufour (2008). Our theoretical survey has revealed that $g$-modes can
indeed be driven in models of Hot DQ stars. However, we found that only
those stars with a sufficiently large amount of He ($X$(He) $>$ 0.25) in
their C-rich envelope mixture could pulsate {\sl in the range of
effective temperature where the real C-atmosphere white dwarfs are
found}. 

\begin{figure}[h]
\begin{center}
\includegraphics[width=30pc]{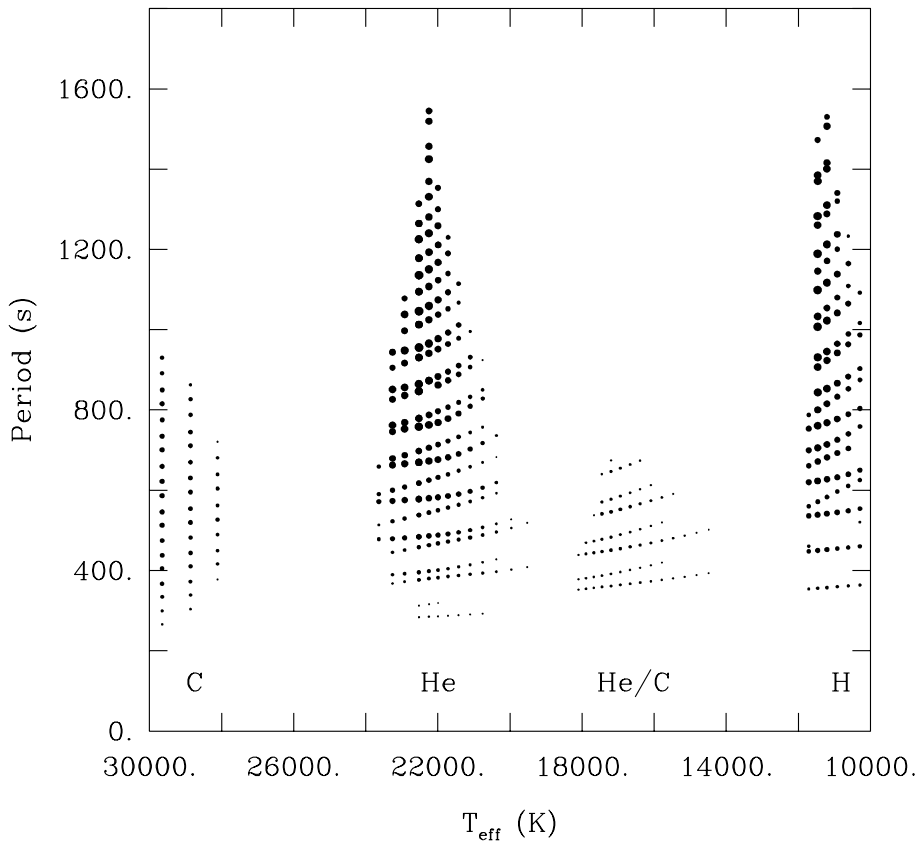}
\end{center}
\caption{Predicted spectra of excited dipole $g$-modes computed
from four distinct evolutionary sequences, each characterized by a total
mass of 0.6 $M_{\rm \odot}$, but with a different envelope composition:
pure C, pure He, $X$(He) = $X$(C) = 0.5, and pure H, from left to
right. The so-called ML2/$\alpha$=0.6 version of the mixing-length
theory was used in these calculations. Each dot gives the period of an
unstable mode, and its size represents a logarithmic measure of the
modulus of the imaginary part $\sigma_I$ of the complex
eigenfrequency. The bigger the dot, the more unstable the mode.} 
\end{figure}

In this connection, Figure 1 displays some revealing results. It depicts 
the locations of theoretical instability strips for evolving 0.6 $M_{\rm
\odot}$ white dwarf models with different envelope compositions. Along 
with the usual V777 Her (pure He) and ZZ Ceti (pure H) instability
strips, one can recognize the red edge of the pulsating pure C envelope
white dwarf models. In fact, the pure C instability strip extends all
the way up to the GW Vir regime as described at length in Quirion et
al. (2007). Given that the true red edge is hotter than the $\sim$28,000
K value found in our survey (and see Fontaine \& Brassard 2008), it follows
that pure  C-atmosphere white dwarfs cannot pulsate in the range of
effective temperature where the Hot DQ stars congregate. This is
contrary to the arguments put forward by Montgomery et al. (2008).

On the other hand, models with a mixed He and C envelope composition can
also pulsate, but in different temperature intervals. For instance,
Figure 1 illustrates a new instability strip between the V777 Her and
the ZZ Ceti domains associated with white dwarf models with a mixed
envelope composition specified by $X$(He) = $X$(C) = 0.5. Naively, one
could have expected to find such a strip in between the pure C and pure
He strips, but structural differences in the mixed envelope composition
models explain why this is not so (see Fontaine et al. 2008 for more
detailed explanations on this). In brief, the survey of Fontaine et
al. (2008) revealed that some Hot DQ white dwarfs can indeed
undergo low-order, low-degree $g$-mode pulsational instabilities,
provided that the surface gravity is larger than average and a
substantial amount of He is present in the C-rich envelope
mixture.

\begin{figure}[h]
\begin{center}
\includegraphics[width=30pc]{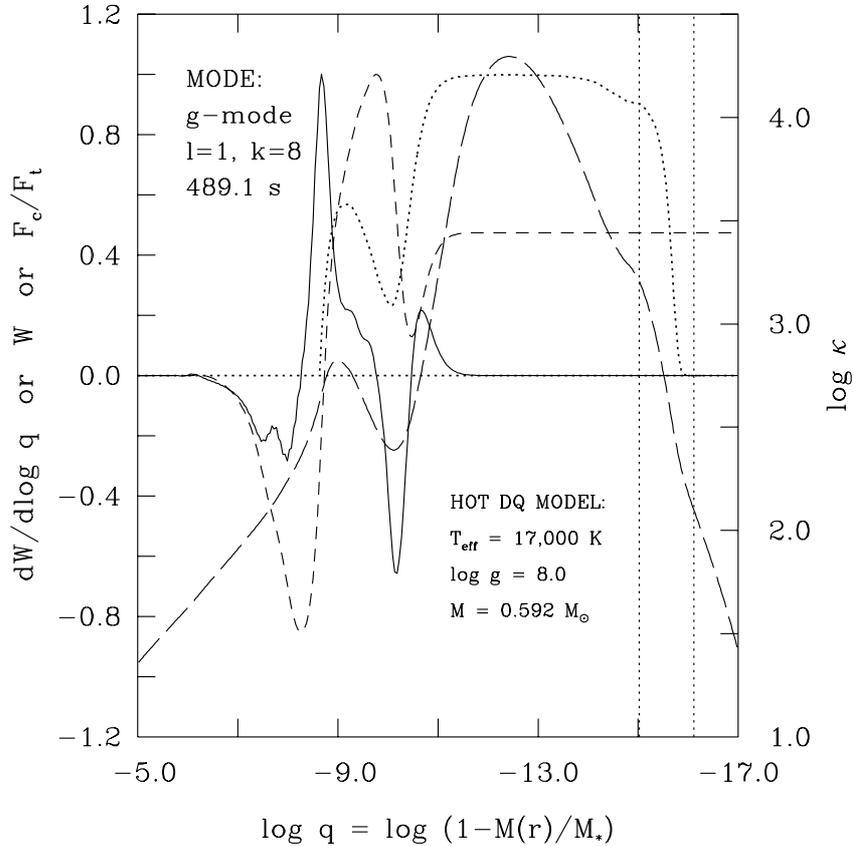}
\end{center}
\caption{Details of the driving/damping process for
  a typical $g$-mode excited in a 17,000 K, ML2 model of a Hot DQ
  white dwarf with a envelope composition $X$(C) =
  $X$(He) = 0.5 and a gravity log $g$ = 8.0. The solid curve shows
  the integrand of the work integral of the mode as a function of
  fractional mass depth. The dashed curve shows the running work
  integral, from left to right, toward the surface of the model. The 
  dotted curve shows the ratio of the convective to total flux. The
  long-dashed curve gives the run of the Rosseland opacity, to be read
  on the RHS ordinate axis. The maximum in the opacity profile, located at
  log $q$ $\simeq$ $-$12.41 and corresponding to a temperature $T$
  $\simeq$ $1.045 \times 10^5$ K, is caused by the partial ionization of
  He II, C III, and C IV in the envelope mixture. The secondary maximum,
  located at log $q$ $\simeq$ $-$8.98 and corresponding to a temperature
  $T$ $\simeq$ $1.177 \times 10^6$ K, is caused by the partial ionization of
  C V, and C VI. The vertical dotted line on the left
  (right) gives the location of the base of the atmosphere at optical
  depth $\tau_R$ = 100 (of the phoptosphere at $\tau_R$ = 2/3).}
\end{figure}

Figure 2 illustrates the details of the driving/damping region in a 
Hot DQ model defined by log $g$ = 8.0, $T_{\rm eff}$ = 17,000 K, and
with an envelope composition specified by $X$(He) = $X$(C) = 0.5. While
this is cooler than the coolest C-atmosphere white dwarf known, there is
a compensation effect related with the gravity such that a similar behavior is
observed in a hotter but higher-gravity model as described in Fontaine
et al. (2008). Out of the many $g$-modes found excited in the retained 
model, we have singled out a representative one with indices $k$ = 8
and $\ell$ = 1. It has a period of 489.1 s. The figure illustrates a
driving/damping situation comparable to the cases of the V777 Her and ZZ
Ceti pulsators, but is more complicated because of the presence of two
maxima in the opacity distribution instead of a single peak in these
other pulsators. In the present case, both opacity maxima (caused by two
distinct partial ionization zones in the envelope mixture) are
``active'' in the sense that they both contribute to the driving/damping
process. What we can observe from the plot is that the regions on the
descending side (going in from the surface) of an opacity bump
contribute locally to driving, while the deeper adjacent zones, where
the opacity plummets to relatively low values, contribute instead to
damping. In the model, the two opacity bumps are relatively close to
each other and are part of a single convection zone. The damping region
between the two bumps is then relatively narrow and the overall
work integral comes out positive, meaning that the mode is globally
excited. 

In parallel with our theoretical investigations, but totally
independently, Montgomery et al. (2008) carried out an observational
search for luminosity variations in six Hot DQ stars accessible to
observations in the winter of 2008. They were able to report the very
exciting discovery that SDSS J1426+5752, one of the Hot DQ's found by
Dufour et al. (2007), pulsates in, at least, one mode with a period of
417 s, thus establishing the existence of a fourth type of pulsating
white dwarf. Full credit should be given to Montgomery et al. (2008) for
this important breakthrough. In addition, they reported that no
luminosity variations were found, to the limit of detection, in the five
other stars in their sample. 

Concerning their work, we have to point out, however, that the
theoretical arguments put forward by Montgomery et al. (2008) to
``predict'' pulsational instabilities in Hot DQ stars are fallacious.  
In particular, the thermal timescale argument that they used is a
necessary but {\sl not} a sufficient condition for instability. Only
full nonadiabatic calculations such as those carried out by Fontaine et
al. (2008), for example, can lead to the final verdict as whether or not a
pulsation mode is unstable in a given stellar model. Montgomery et
al. (2008) did not carry out nonadiabatic calculations and could not,
therefore, conclude about the stability of Hot DQ models. Fortunately,
this did not prevent them from going to the telescope and discovering the
first pulsating C-atmosphere white dwarf. In the end, this is really
what matters.

\section{Follow-up photometric observations of SDSS J1426+5752}

\begin{figure}[h]
\begin{center}
\includegraphics[width=25pc,angle=-90]{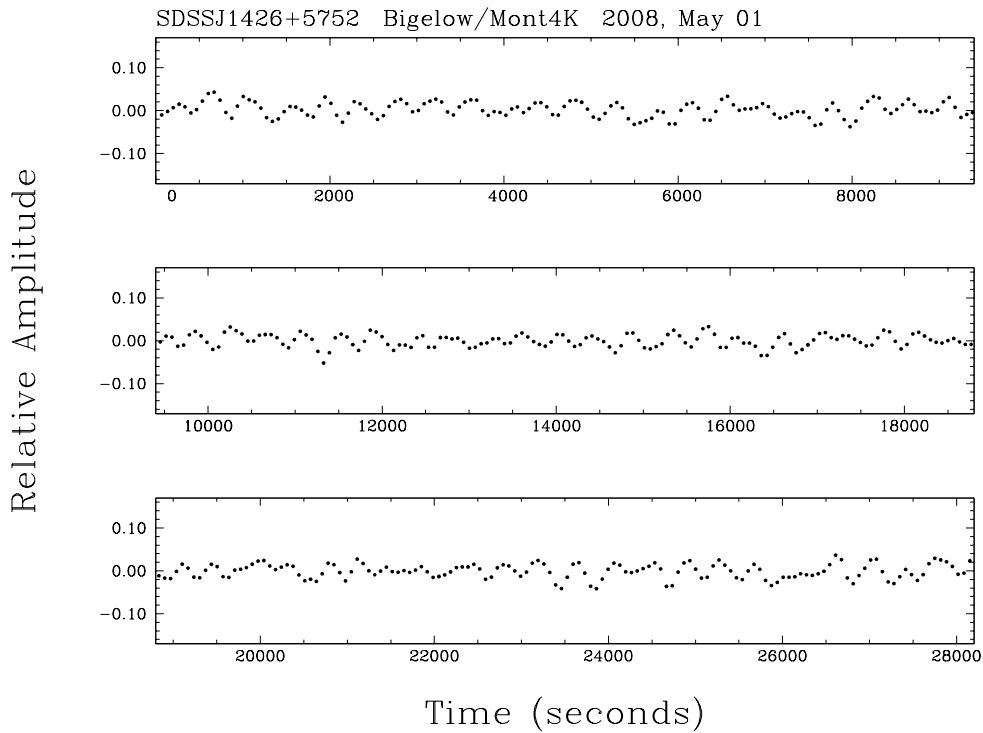}
\end{center}
\caption{Sample broadband light curve of SDSS J1426+5752
obtained with the Mont4K CCD camera mounted on the Steward Observatory
1.6-m Kuiper telescope through a Schott 8612 filter. Each plotted point
corresponds to an effective sampling time of 67 s.}
\end{figure}

Following this discovery, Green, Dufour, \& Fontaine (2008, in
preparation) undertook follow-up wide band photometric observations of
SDSS J1426+5752 at the Steward Observatory 1.6-m telescope on Mount
Bigelow with the help of the new Montr\'eal 4K$\times$4K CCD camera
(Mont4K), a joint venture between the University of Arizona and the
Universit\'e de Montr\'eal. Some 106 h of observations were obtained on
this rather faint star ($g$ = 19.2). Figure 3 illustrates one of the
nightly light curves obtained by Green et al., leaving no doubt as to
the variability of SDSS J1426+5752. The Fourier amplitude spectrum for
the full data set confirms the presence of a dominant pulsation with a
period of 417 s and of its first harmonic  as first reported by Montgomery
et al. (2008). In addition, it also reveals the likely presence of an
additional pulsation with a period of 319 s at the 4.9 sigma level. 
Hence, with at least two independent periodicities uncovered, SDSS
J1426+5752 can be considered as a multiperiodic pulsator like the other
types of pulsating white dwarfs. 

\section{Follow-up spectroscopic observations of SDSS J1426+5752}

\begin{figure}[h]
\begin{center}
\includegraphics[width=30pc]{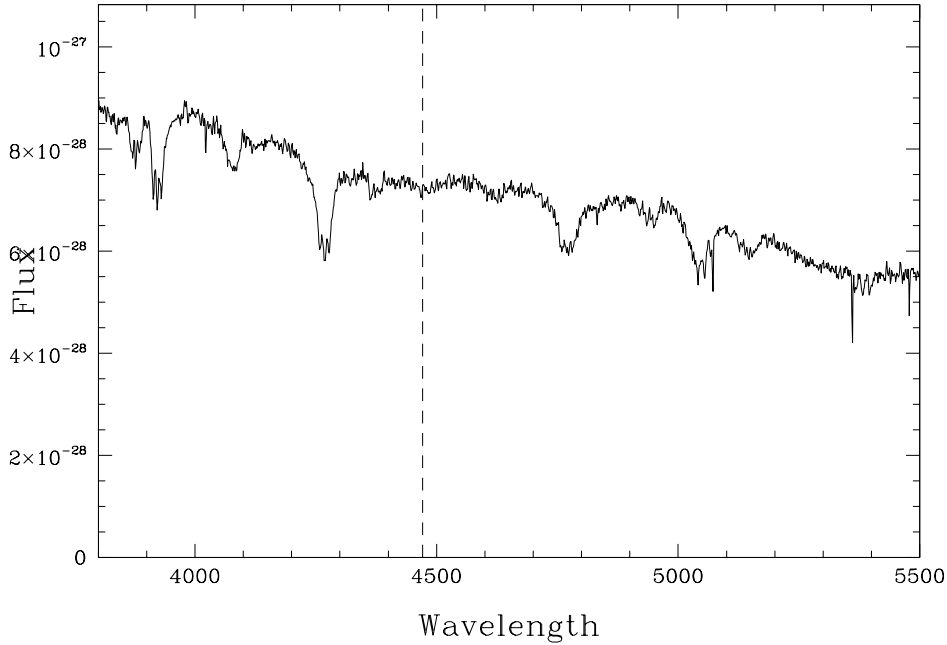}
\end{center}
\caption{Optical spectrum of SDSS J1426+5752 obtained at the Multiple
  Mirror Telescope. The strong carbon lines are obvioulsy affected by
  Zeeman splitting due to the presence of a strong magnetic field. The
  same is true of the faint HeI 4471 line. Although relatively weak,
  that line implies an atmospheric helium abundance comparable to that
  of carbon in that star.} 
\end{figure}

In view of the rather poor SDSS spectrum available for SDSS J1426+5752,
follow-up spectroscopic observations were pursued by Dufour et al. (2008b) 
using both the MMT and one of the Keck telescopes. The objective
was, firstly, to obtain a sufficiently good spectrum for detailed
atmospheric modeling and, secondly, to search for the presence of He
required to account for the observed pulsational instabilities according
to the nonadiabatic calculations of Fontaine et al. (2008). The spectral
analysis of the improved spectra readily revealed the presence of a
substantial amount of helium in the atmosphere of SDSS J1426+5752, an
abundance comparable to that of carbon by mass fraction. This is in line
with the expectations of nonadiabatic pulsation theory which require an
important He ``pollution'' in the atmosphere/envelope of SDSS J1426+5752
for it to pulsate at its current effective temperature. In this context,
Dufour et al. (2008a) also showed that the five other objects found not
to vary by Montgomery et al. (2008) are, contrary to SDSS J1426+5752,
not expected to vary.

To add to this small success, an unexpected surprise came out of the
follow-up spectroscopic observations of Dufour et al. (2008b). It was
found that the strong carbon lines seen in the spectrum of SDSS
J1426+5752 feature Zeeman splitting, a structure that could not
be seen in the original noisy SDSS spectrum. Figure 4 illustrates our
MMT spectrum. The observed splitting between the $\pi$ and $\sigma$
components implies a large scale magnetic field of about 1.2 MG. Hence,
SDSS J1426+5752 is both a pulsating and a magnetic white dwarf. As there
is no sign of binarity in either the light curve or the optical spectrum
(although the phase coverage has been quite limited), SDSS J1426+5752 is
most likely the first example of an isolated pulsating white dwarf with
a large detectable magnetic field. As such, it is the white dwarf
equivalent of an roAp star. Interestingly, SDSS J1426+5752 could have
been itself an roAp star in its distant past.

\section{Conclusion}

It remains to be seen if other stars similar to SDSS J1426+5752 will be
found or if it will remain an isolated ``freak''. If we adopt a
conservative point of view, it takes at least {\sl two} members to
define a ``class'', so we should perhaps refrain from referring to it as the
prototype of a class for the time being. Note, however, that at the time
of writing, rumors have it that two more Hot DQ stars may also be
pulsating objects. Be it as it may, SDSS J1426+5752 is certainly
different from the other kinds of pulsating white dwarfs that we know of
(GW Vir, V777 Her, and ZZ Ceti). It is hoped that further data releases
from the SDSS project might reveal siblings of this fascinating star.
A fourth asteroseismological window has thus been opened through which
one can further study the properties of white dwarf stars. 

\section*{References}
\begin{thereferences}
\item Dufour P, Fontaine G, Liebert J, Schmidt G D and
Behara N 2008a {\sl ApJ} {\bf 683} 978
\item Dufour P, Fontaine G, Liebert J, Williams K A and
Lai D K 2008b {\sl ApJ} {\bf 683} L167
\item Dufour P, Liebert J, Fontaine G and Behara N 2007
  {\sl Nature} {\bf 450} 522
\item Fontaine G and Brassard P 2008, {\sl PASP} {\bf 120} 1043 
\item Fontaine G, Brassard P and Dufour P 2008{\sl ApJ} {\bf
  483} L1
\item Fontaine G and Van Horn H M 1976 {\sl ApJS} {\bf 31} 467
\item Montgomery M H, Williams K A, Winget D E, Dufour P,
DeGennaro S and Liebert J 2008 {\sl ApJ} {\bf 678} L51
\item Quirion P-O, Fontaine G and Brassard P 2007 {\sl ApJS} {\bf 171} 219
\end{thereferences}

\end{document}